\documentclass[pra,showpacs,twocolumn]{revtex4}  
\usepackage{revsymb}
\usepackage[cp1250]{inputenc}   
\usepackage[T1]{fontenc}
\usepackage{amssymb,amsmath,amscd,amsfonts}
\usepackage[dvips]{graphicx}
\usepackage{indentfirst}
\usepackage{multirow}
\usepackage{color}
\usepackage{hyperref}
\usepackage{dsfont,lscape}

\newcommand{\ket}[1]{|#1\rangle}

\newcommand{\bra}[1]{\langle#1|}
\newcommand{\braket}[2]{\langle #1 | #2 \rangle}
\renewcommand{\t}[1]{\textrm{#1}}

\newcommand{\Tr}{\t{Tr}}
\newcommand{\mat}[2]{
\left(\begin{array}{#1}
#2
\end{array}
\right)}

\renewcommand{\eqref}[1]{Eq.~(\ref{#1})}
\newcommand{\op}[1]{\boldsymbol{#1}}
\newcommand{\su}[1]{{SU(#1)}}
\newcommand{\tr}{\text{Tr}}

\newcommand{\param}{\op{\theta}} 

\newcommand{\tabref}[1]{Tab.~\ref{#1}}
\newcommand{\figref}[1]{Fig.~\ref{#1}}

\begin{document}

\title{Optimal state for keeping reference frames aligned and the Platonic solids}

\author{Piotr Kolenderski}
\homepage{http://www.fizyka.umk.pl/~kolenderski}

\affiliation{Institute of
Physics, Nicolaus Copernicus University, ul. Grudzi{a}dzka 5, 87-100
Toru{\'n}, Poland}

\author{Rafal Demkowicz-Dobrzanski}
\affiliation{Institute of
Physics, Nicolaus Copernicus University, ul. Grudzi{a}dzka 5, 87-100
Toru{\'n}, Poland}

\date{\today}

\begin{abstract}
The optimal $N$ qubit states featuring highest sensitivity to small misalignment of cartesian reference frames are found using the Quantum Cram\'{e}r-Rao bound. It is shown that the optimal states are supported on the symmetric subspace and hence are mathematically equivalent to a single spin $J=N/2$. Majorana representation of spin states is used to reveal a beautiful connection between the states optimal for aligning reference frames and the platonic solids.
\end{abstract}

\maketitle

\section{Introduction}
A meaningful communication requires a common frame of reference between the sender and the receiver. Depending on the nature of transmitted information, different types of references are needed e.g. a common time reference, phase reference, direction reference, cartesian frame of reference etc. Establishing a common frame can be achieved by sending a physical system prepared in a proper state e.g. a clock, a gyroscope, which allows two parties for aligning, and calibrating their devices accordingly.

The quality of the alignment depends on the quality in which the relevant information is encoded and decoded from the physical system. Roughly speaking, the higher is the relevant signal to noise ratio in the system the better is the alignment. When looking for fundamental limitations on the precision of alignment, quantum mechanics provides ultimate bounds on the signal to noise ratio arising from unavoidable uncertainty inherent in any quantum measurement. The problem of optimal alignment is thus the following: under given physically motivated restrictions, e.g. finite number of particles, limited energy, etc.  find the state and the measurement that yield alignment with the highest accuracy possible.

Mathematically the problem may be formulated as follows. Consider a quantum system (in particular the system may consist of many copies of smaller subsystems) and let $\ket{\psi}$ be a quantum state prepared by one of the parties. After being sent to the second party, which uses a different reference frame, the evolution of the state may be described by a unitary representation $U_g$, where $g \in \mathcal{G}$ is an element of a group describing possible transformation of reference frames, e.g.  $\mathcal{G}=U(1)$  (phase), $\mathcal{G}=SU(2)/U(1)$ (direction), $\mathcal{G}=SU(2)$ (cartesian frame). The transmitted state $\ket{\psi_g}=U_g \ket{\psi}$ is subsequently measured be the second party in order to establish the relation between its own reference frame and that of the first party. The measurement is described by a set of positive operators $\Pi_\xi \geq 0$, $\sum_\xi \Pi_\xi = \openone$. Depending on the result obtained, a guess $\tilde{g}_\xi$ (the estimator function) is made concerning the true value $g$
of the group element responsible for the transformation of the state. Finally a cost function $C(g,\tilde{g})$ should be defined penalizing for inaccurate estimation of $g$. The problem is solved once the state and the measurement is found minimizing the average value of the cost function (for a review see e.g. \cite{Bartlett2007}). From a physical point of view any reasonable cost function should be both left invariant: $C(h g, h g_\xi)=C(g,g_\xi)$ which makes
this function sensitive only to relative transformation of the state, and right invariant $C(g h, g_\xi  h)=C(g,g_\xi)$ which can be seen as independence from background reference frame transformations. It can be shown \cite{Chiribella2005, Bartlett2007} that such a cost function can always be written as
\begin{equation}
\label{eq:invcostfunction}
    C(g,g_\xi)=\sum_q c_q \chi_q(g_\xi^{-1} g),
\end{equation}
where $c_q$ are arbitrary coefficients and $\chi_q$ are characters of the group $\mathcal{G}$.

\begin{table*}[t]
\begin{center}
    \begin{tabular}{|c||c|c|c||c|c|c|}
        \hline
        & \multicolumn{3}{c||}{global approach} &  \multicolumn{3}{c|}{local approach}\\
        \cline{2-7}
        & phase & direction & cartesian &  phase & direction & cartesian \\
        \hline\hline
         $N$ qubits  &\multirow{2}{*}{$\sum_{m=-J}^J \alpha_m \ket{J,m}$,}\cite{berry2000} &
         $\sum_{j=0}^{J}\beta_j \ket{j,0}$, \cite{bagan2001}  &
         $\ket{J,J}+ \sum_{j=0}^{J-1}\gamma_j \ket{e_j}$, \cite{Chiribella2004} &
         \multirow{2}{*}{$\ket{J,J}+\ket{J,-J}$,} \cite{bollinger1996}  & \multirow{2}{*}{$\ket{J,0}$} &
         \multirow{2}{*}{\t{see Table~\ref{tab:OptimalStates}.}} \\
        \cline{1-1}\cline{3-4}
        spin $J=N/2$& & $\ket{J,J}$, \cite{holevo1982}& $\ket{J,J}$, \cite{holevo1982} & & & \\
        \hline
    \end{tabular}
    \end{center}
    \caption{
Summary of known results in the field of optimal alignment of reference frames.
    The table presents the optimal unnormalized states for encoding phase, direction and cartesian frame of reference,
    where $\alpha_m$, $\beta_j$, $\gamma_j$ are nonzero and they specific values depend on the cost function chosen.
    For simplicity we assumed $N$ is even. When writing states of $N$ qubits we use the notation corresponding to decomposition \eqref{eq:Decomposition}.
    When writing states of $N$ qubits we use the notation which is in agreement with the following
    decomposition of $N$ fold tensor product of single qubit Hilbert space:
    $\mathcal{H}^{\otimes N}= \bigoplus_{j=0}^{N/2} \mathcal{H}_j \otimes \mathbb{C}_{d_j}$, where $\mathcal{H}_j$ carries $2j+1$ dimensional irreducible representation of $SU(2)$ while $\mathbb{C}_{d_j}$ is $d_j$ dimensional multiplicity space reflecting the existence of multiple equivalent representation with the same $j$.
The state $\ket{e_j}$ is an maximally entangled state in $\mathcal{H}_j \otimes \mathbb{C}_{d_j}$, while by $\ket{j,m}$ we mean a product state $\ket{j,m} \otimes \ket{\xi} \in \mathcal{H}_j \otimes \mathbb{C}_{d_j}$, where the form of the state $\ket{\xi}$ does not play any role.
In the local approach there is no advantage in using $N$ distinguishable qubits
    to using $N$ qubits in a fully symmetric state or equivalently a single spin $J=N/2$.
    This paper presents the optimal states for cartesian frame alignment in the local approach (see \tabref{tab:OptimalStates}).
} \label{tab:states}
\end{table*}

Two different approaches to estimation problem are most often pursued. (i) In \emph{the global approach} one assumes the complete ignorance about the actual value of $g$ --
in mathematical terms one assumes an a priori probability distribution $p(g)$ to be uniform with respect to the Haar measure on $G$.
The quantity to be minimized is:
\begin{equation}
    \bar{C}(g) = \int \t{d}\mathcal{G} \sum_\xi \bra{\psi_g} \Pi_\xi \ket{\psi_g} C(g,\tilde{g}_\xi),
\end{equation}
where $\t{d}\mathcal{G}$ denotes the normalized Haar measure.
(ii) In \emph{the local approach}  one assumes that the group element is close to the known value $g$ and the goal is to find a measurement and the estimator featuring the highest sensitivity to small variations of $g$. Let $g(\param)$ be a parametrization of the group with $p$ real parameters $\param=(\theta_1,\dots,\theta_p)$. Since we consider only small variations of group elements, a cost function can be well approximated by a quadratic form $G$:
$$C(g(\tilde{\param}),g(\param)) \approx (\param-\op{\tilde{\theta}})^T \op {G} (\param-\op{\tilde{\theta}}).$$
We need to find a state $\ket{\psi}$, a measurement $\Pi_\xi$ and an estimator $\tilde{\param}_\xi$
minimizing:
\begin{equation}
    \label{eq:costqform}
    \bar{C}(g(\param)) = \sum_\xi \bra{\psi_{\param}} \Pi_\xi
    \ket{\psi_{\param}} (\param-\op{\tilde{\theta}}_\xi)^T
    \op G (\param-\op{\tilde{\theta}}_\xi),
\end{equation}
where $\ket{\psi_{\param}}= \ket{\psi_{g(\param)}}$ (in what follows we will write $\param$ instead of $g(\op{\theta})$ to simplify the notation). Fortunately, in the local approach the quantum Cram\'{e}r-Rao inequality \cite{helstrom1976} gives a lower bound on the minimal cost without the need of optimizing over estimators and measurements. For an arbitrary measurement and any unbiased estimator (i.e the one which average value over measurement outcomes yields the true value) the cost function is bounded by:
\begin{equation}
\label{eq:cramerrao}
    \bar{C}(\param) \geq {\Tr(\op{G} \op{F}^{-1})}/{n},
\end{equation}
where $n$ is the number of repetitions of an experiment and $\op{F}$ is the Fisher information matrix, which for the case of pure states can be explicitly written as:
\begin{equation}\label{eq:FisherMatrixDef}
    \op{F}_{ij}= 4\t{Re}(\braket{\psi_{\param,i}}{\psi_{\param,j}} -
     \braket{\psi_{\param}}{\psi_{\param,i}} \braket{\psi_{\param,j}}{\psi_{\param}}),
\end{equation}
where $\ket{\psi_{\param,i}}=\frac{\partial \ket{\psi_{\param}}}{\partial \theta_i}$. For the single parameter estimation ($p=1$) the Cram\'{e}r-Rao bound is tight \cite{helstrom1976} at least asymptotically for $n \rightarrow \infty$.
In the multiparameter case, due to potential non-commutativity of optimal measurements for different parameters, the bound is not always achievable. In the case of pure states, however, the bound is indeed tight provided the following condition holds \cite{matsumoto2002} $\t{Im} \braket{\psi_{\param,i}}{\psi_{\param,j}}=0$.
As will be seen later in the paper, in the problem of the cartesian reference frame alignment, the states minimizing right hand side in \eqref{eq:cramerrao} indeed satisfy the above condition. This justifies the use of Cram\'{e}r-Rao inequality as a tool for looking for the optimal states for
cartesian reference frame alignment in the local approach.

Let us first briefly remind the known results in quantum estimation theory concerning the reference frames alignment. The two most commonly used physical systems for this purpose are: the system of $N$ distinguishable qubits (physically equivalent to $N$ spins $1/2$ or polarization states of $N$ photons traveling in separate time-bins), and the system of $N$ qubits in a fully symmetric state (bosonic states, e.g. polarization states of photons traveling in a single time-bin).
\tabref{tab:states} lists the optimal states for phase, direction and cartesian frame encodings, both when using $N$ distinguishable qubits
and when using them in a fully symmetric state or equivalently using a single spin $J=N/2$.

This paper fills the missing gap by presenting the optimal state for reference frame alignment in the local approach \footnote{In a different paradigm, in which it is allowed to use an extra ancillary system of arbitrary dimension as a reference \cite{ballester2004}, the optimal states for aligning cartesian reference frame in the local approach were found in \cite{ballester2005}.} and gives an intuitive picture of the states using Majorana representation \cite{Majorana1932}. Moreover we prove that in the local approach distinguishability of the qubits gives no advantage over the use of $N$ qubits in the fully symmetric state or a single spin $J=N/2$ -- a fact known for phase and direction reference alignment.

\section{Optimization}
Let us consider the most general pure state
$\ket{\psi}\in \mathcal{H}^{\otimes N}$ of $N$ distinguishable qubits.
Using a decomposition of
\begin{equation}\label{eq:Decomposition}
  \mathcal{H}^{\otimes N}= \bigoplus_{j=0}^{N/2} \mathcal{H}_j \otimes \mathbb{C}_{d_j},
\end{equation}
where $\mathcal{H}_j$ carries $2j+1$ dimensional irreducible representation of $SU(2)$ while $\mathbb{C}_{d_j}$ is $d_j$ dimensional multiplicity space reflecting the existence of multiple equivalent representations with the same $j$, we can write the state as
$\ket{\psi}=\sum_{j=0}^{N/2} \alpha_j \ket{\psi^{(j)}}$
where $\ket{\psi^{(j)}}$ is an arbitrary state in $\mathcal{H}_j \otimes \mathbb{C}_{d_j}$
(for $N$ odd the summation starts from $j=1/2$). Transmission of the state to a different cartesian frame yields the output state of the form: $\ket{\psi_{\param}} = U_{\param} \ket{\psi}$,
where $U_{\param}$ is the tensor representation of $\su{2}$ acting on $\mathcal{H}^{\otimes N}$.
Choosing a convenient parametrization of the $\su{2}$ group $g = \exp(i \param \op{J})$
 the output state reads:
\begin{equation}\label{eq:statebob}
\ket{\psi_{\param}} =   \sum_{j=0}^{N/2} \alpha_j \exp(i \param \op{J^{(j)}}) \otimes  \openone_{d_j} \ \ket{\psi^{(j)}} ,
\end{equation}
where $\op{J}^{(j)}=(\op{J}^{(j)}_x,\op{J}^{(j)}_y,\op{J}^{(j)}_z)$ is the angular momentum operator in the spin $j$ representation. Without losing generality we may assume that we study sensitivity of reference frame alignment around $\param = \op{0}$.

According to \eqref{eq:invcostfunction}, when estimated $\tilde{\param}$ differs from the true $\param = \op{0}$,
the general cost function reads
$$C(\tilde{\param})=\sum_j c_j \Tr[\exp(i \tilde{\param} \op{J}^{(j)})].$$
Making use of the fact that angular momentum operators are traceless and orthogonal in the Hilbert-Schmidt metric, expansion of  the cost function up to the second order yields
$C(\tilde{\param}) \propto \|\tilde \param\|^2$,
where the proportionality constant depends on $c_j$ coefficients. Hence the quadratic form $\op{G}$ (see \eqref{eq:costqform}) is proportional to identity in this parametrization.

According to the Cram\'{e}r-Rao bound [\eqref{eq:cramerrao}], optimal states are the ones for which $\tr \op{G}\op{F}^{-1}$ is minimal. Since $\op G \propto \openone$, the problem simplifies to finding the states  which  minimize $\tr \op{F}^{-1}$. Using \eqref{eq:statebob} the Fisher matrix [\eqref{eq:FisherMatrixDef}] in our parametrization reads:
\begin{equation}\label{eq:FisherMatrix}
      \op{F}_{ik}=4 \left(\frac{1}{2}\bra{\psi}J_i J_k+J_k J_i\ket{\psi}-\bra{\psi}J_i\ket{\psi}\bra{\psi} J_k\ket{\psi}\right),
\end{equation}
where $i,k=1,2,3$ and $\op{J} = \oplus_{j=0}^{N/2} \op{J}^{(j)} \otimes \openone$.
Note that the Fisher matrix $\op{F}$ is proportional to the spin covariance matrix for the state $\ket{\psi}$. The problem of minimization of the average cost function is thus equivalent to
minimization of the sum of inverse eigenvalues of the spin covariance matrix.

Since the harmonic mean is always smaller that the arithmetic mean we get the following inequality:
\begin{equation}
    \label{eq:harmarith}
    \tr \op{F}^{-1} \geq {9}/{\tr \op{F}},
\end{equation}
with equality if and only if $\op{F} \propto \openone$. Notice that $\tr \op{F} = \sum_i  \bra{\psi} J_i^2 \ket{\psi} - \bra{\psi} J_i \ket{\psi}^2$,
and is thus bounded
\begin{equation}
    \label{eq:jbound}
    \tr \op{F} \leq \sum_{j=0}^{N/2} |\alpha_j|^2 j(j+1) \leq  J(J+1),
\end{equation}
where $J=N/2$. States which yield equality in the above equation with equal variances of all
$J_i$ operators guarantee equality in \eqref{eq:harmarith} and thus are optimal.
Hence, we need to find states with variances $\Delta J_i=J(J+1)/3=N(N+2)/12$. Notice that if such states exist they have to be states from the fully symmetric subspace $j=N/2$. If we find such states, we simultaneously prove that using other subspaces with $j<N/2$ is not helpful, and hence distinguishability of qubits is useless.

\section{Majorana representation}
To look for the optimal states in the fully symmetric subspace, we will make use of the Majorana representation, which provides a beautiful geometrical picture of symmetric states. Every symmetric state $\ket{\psi}$ can be
written as a symmetrized product state \cite{Majorana1932, Prenrose1984}:
\begin{equation}
\ket{\psi}= \mathcal{N} \sum_{\sigma \in S_N} \ket{\vec{n}_{\sigma(1)}} \otimes \ket{\vec{n}_{\sigma(2)}} \otimes \dots \otimes
\ket{\vec{n}_{\sigma(N)}},
\end{equation}
where the summation is performed over all permutations of $N$ elements, $\mathcal{N}$ is a normalization factor, and
$\ket{\vec{n}}$ is a single qubit state with Bloch vector pointing in $\vec{n}$ direction.
The correspondence between symmetric states and product states is one to one,
and as a result we may represent a symmetric state uniquely as $N$ points on the Bloch sphere. This is the Majorana representation
of a symmetric state.
Obtaining a symmetric state from a product state is straightforward, going in the reverse direction is less trivial and one has to proceed as follows. Instead of a Bloch vector $\vec{n}=[\sin\theta \cos\phi, \sin\theta\sin\phi,\cos\theta]$, we may use a stereographic projection and
parameterize a single qubit state with a single complex number $z=e^{-i \phi}\cot{\theta/2}$, where $z= \infty$ for $\theta=0$. Let $\ket{z_{\perp}}$ be a state orthogonal to $\ket{z}$. For a given state
$$\ket{\psi}=\sum_{m=-N/2}^{N/2} a_m \ket{N/2,m}$$
the overlap $\bra{z_\perp}^{\otimes N}\ket{\psi}$, 
up to an irrelevant function of $z$ having no roots, is proportional to the {\em Majorana polynomial}:
\begin{equation}\label{eq:MajoranaPolymnomial}
    \sum_{m=-N/2}^{N/2}(-1)^{k} \mat{c}{N \\ \frac{N}{2}+m}^{\frac{1}{2}} a_m  z^{\frac{N}{2}+m} 
\end{equation}
By the fundamental theorem of algebra, every polynomial can be uniquely factored. Thus for each symmetric state there exist a unique set of $N$ complex numbers composed of $\tilde{N}$ roots of the Majorana polynomial $\{z_1,z_2,\dots,z_{\tilde N} \}$ supplemented by $N-\tilde N$ element set of $\infty$ that yields the product state $\ket{{z}_1} \otimes \ket{{z}_2} \otimes \dots \otimes \ket{{z}_N}$.

Note that under the action of \su{2} group on a symmetric state the corresponding points in the Majorana representation are rotated as a rigid solid. Hence, when looking for the optimal state for reference frame alignment in the Majorana representation we intuitively should look for set of $N$ points on the Bloch sphere which are ``the most sensitive'' to arbitrary rotation. What comes naturally to mind, is to take as Majorana points vertices of the most symmetric solids i.e. the platonic solids. Amazingly, all five Platonic solids -- tetrahedron $N=4$, octahedron $N=6$, cube $N=8$, icosahedron $N=10$, dodecahedron $N=20$ yield variances $\Delta J_i=N(N+2)/12$ (this was observed in \cite{Zimba2006} and the corresponding states were named anti-coherent). Hence in the face of previous discussion the anti-coherent states are optimal for cartesian reference frame alignment. Analyzing the geometrical structure of the optimal states obtained from Platonic solids one can easily generalize for other values of $N$. Table \ref{tab:OptimalStates} presents states written as a simple generalization of the platonic solids states, which provide optimal states for reference frame alignment for every
even $N\geq 4$. Notice that for some $N$ you may find optimal states in many different classes. The state corresponding to the lowest allowed $N$ in each class corresponds to the perfect solid state.

\begin{table}[ht]
  \centering
  \begin{tabular}{|c|c|}
    \hline
    class name & state $\ket{\psi}$\\
    \hline
    \hline
    \multirow{2}{*}{tetrahedron}     & $ \sqrt{\frac{N+2}{4 N+2}}\ket{-N/2}+ \sqrt{\frac{3N}{4 N+2}}\ket{\frac{N+2}{6}}$ \\
           &\t{for }   $N \mod 6 = 4,\quad N\geq 4$ \\
    \hline
    \multirow{2}{*}{octahedron}  & $ \frac{1}{\sqrt{2}}\ket{-\sqrt{\frac{N(N+2)}{12}}}+
    \frac{1}{\sqrt{2}}\ket{\sqrt{\frac{N(N+2)}{12}}}$ \\
    & \t{for }   $\sqrt{\frac{N(N+2)}{12}}\in \mathds{N},\quad N\geq 6$\\
    \hline
    \multirow{2}{*}{cube}  & ${\sqrt{\frac{N+2}{6N}}}\ket{\frac{-N}{2}}+{\sqrt{\frac{N-1}{N}}}\ket{0}+{\sqrt{\frac{N+2}{6N}}}\ket{\frac{N}{2}}$\\
     & \t{for }$N\geq 8$ \\
    \hline
    \multirow{2}{*}{icosahedron }  & $\alpha\ket{-\frac{N}{2}+1}+\sqrt{1-\alpha^2}\ket{0}+\alpha\ket{\frac{N}{2}-1}$\\
     & \t{for } $\alpha=\sqrt{\frac{N (N+2)}{6 (N-1)^2}}\quad N\geq 10$\\
    \hline
    \multirow{3}{*}{dodecahedron}      & $\frac{1}{2}\ket{-\frac{N}{2}}+\alpha\ket{-\frac{N}{4}}+\beta\ket{0}+\alpha\ket{\frac{N}{4}}+\frac{1}{2}\ket{\frac{N}{2}} $ \\
     &  for $\alpha=\sqrt{{2}/{N}}$, $\beta=\sqrt{{N-8}/{2N}}$\\
           & $N\mod 4=0,\quad N\geq 20$ \\
    \hline
  \end{tabular}
  \caption{The states for even number $N$ of qubits derived from Platonic solid state, optimal for local estimation of reference frame.}
  \label{tab:OptimalStates}
\end{table}

\begin{figure}[h]
\includegraphics[width=0.85\columnwidth]{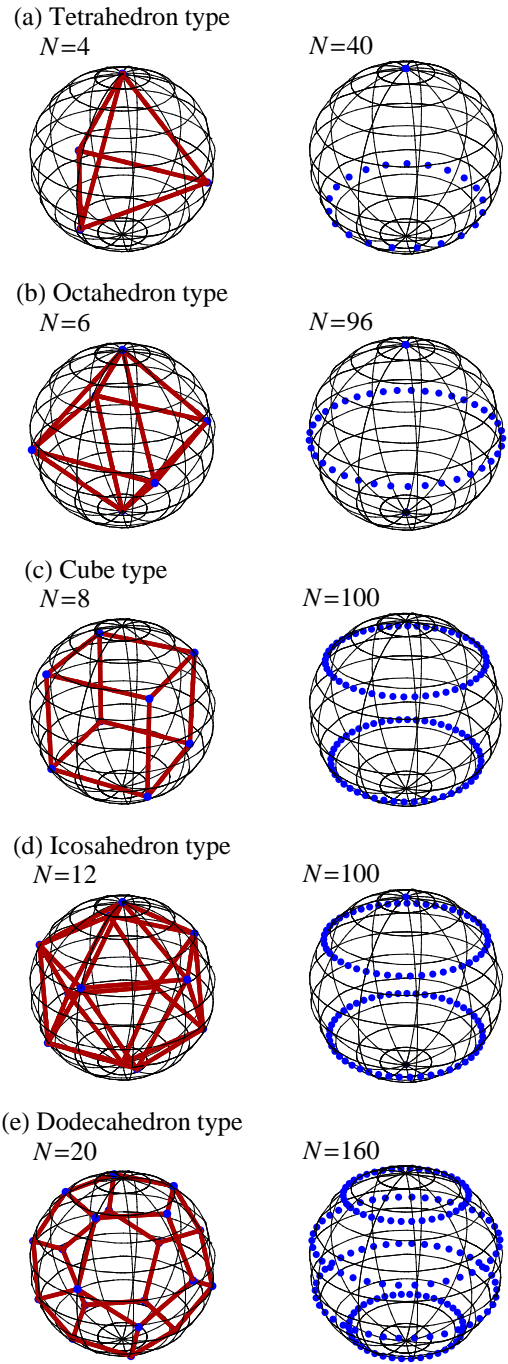}
\caption{(Color online) Majorana representation of Platonic solid classes of optimal states for local estimation of reference frame.  Points on the poles in the right side pictures of a) and b) are degenerate.}\label{fig:Majorana}
\end{figure}

The Majorana representation of state presented in \tabref{tab:OptimalStates} are shown in \figref{fig:Majorana}. We do not claim to present all classes of optimal state for reference frame alignment and it seems from numerical calculations that they are infinitely many of them. This is completely different from direction reference alignment case where $\ket{J,0}$ is the only optimal state.
For odd $N$ analogous states perform optimally, e.g. in \emph{cube}, \emph{icosahedron}, \emph{dodecahedron} classes one only needs to replace $\ket{0}$ state with  $\ket{1/2}$ or $\ket{-1/2}$  state and adjust weights to make the covariance matrix diagonal with $\Delta J_i=N(N+2)/12$.

We should also mention that for and only for $N=1$, $N=2$, $N=3$, $N=5$
there are no pure states yielding a diagonal covariance matrix with $\Delta J_i=N(N+2)/12$.
In these cases the optimal states have to be found by directly maximizing the sum of inverse eigenvalues of the Fisher information matrix, which can be easily done numerically. 

\section{Conclusions}
In this paper we have found the optimal states for aligning cartesian reference frames in the approach where deviations from perfect alignment are small.
Using Cram\'{e}r-Rao bound we have shown that the problem simplifies to the search for $N$ qubit states with diagonal spin covariance matrix with maximal possible variances of all three spin components $\Delta J_i = N(N+2)/12$. We have used the Majorana representation where Platonic solids correspond to the optimal states, and proposed classes of states optimal for arbitrary high number of qubits $N$. As a byproduct we have also proven that distinguishability of qubits is useless for cartesian reference frame alignment in the local approach.

\section{Acknowledgements}
We acknowledge fruitful discussions and great support from Konrad Banaszek, and
a discussion with Paolo Perinotti. This work has been supported by the Polish budget funds for
scientific research projects in years 2005-2008 (Grant No. 1 P03B 011 29) and the European Commission under the Integrated Project Qubit Applications (QAP) funded by the IST directorate as Contract Number 015848.


\begin{thebibliography}{14}
\expandafter\ifx\csname natexlab\endcsname\relax\def\natexlab#1{#1}\fi
\expandafter\ifx\csname bibnamefont\endcsname\relax
  \def\bibnamefont#1{#1}\fi
\expandafter\ifx\csname bibfnamefont\endcsname\relax
  \def\bibfnamefont#1{#1}\fi
\expandafter\ifx\csname citenamefont\endcsname\relax
  \def\citenamefont#1{#1}\fi
\expandafter\ifx\csname url\endcsname\relax
  \def\url#1{\texttt{#1}}\fi
\expandafter\ifx\csname urlprefix\endcsname\relax\def\urlprefix{URL }\fi
\providecommand{\bibinfo}[2]{#2}
\providecommand{\eprint}[2][]{\url{#2}}

\bibitem[{\citenamefont{Bartlett et~al.}(2007)\citenamefont{Bartlett, Rudolph,
  and Spekkens}}]{Bartlett2007}
\bibinfo{author}{\bibfnamefont{S.~D.} \bibnamefont{Bartlett}},
  \bibinfo{author}{\bibfnamefont{T.}~\bibnamefont{Rudolph}}, \bibnamefont{and}
  \bibinfo{author}{\bibfnamefont{R.~W.} \bibnamefont{Spekkens}},
  \bibinfo{journal}{Rev. Mod. Phys.} \textbf{\bibinfo{volume}{79}},
  \bibinfo{eid}{555} (\bibinfo{year}{2007}).

\bibitem[{\citenamefont{Chiribella et~al.}(2005)\citenamefont{Chiribella,
  D'Ariano, and Sacchi}}]{Chiribella2005}
\bibinfo{author}{\bibfnamefont{G.}~\bibnamefont{Chiribella}},
  \bibinfo{author}{\bibfnamefont{G.~M.} \bibnamefont{D'Ariano}},
  \bibnamefont{and} \bibinfo{author}{\bibfnamefont{M.~F.}
  \bibnamefont{Sacchi}}, \bibinfo{journal}{Phys. Rev. A}
  \textbf{\bibinfo{volume}{72}}, \bibinfo{pages}{042338}
  (\bibinfo{year}{2005}).

\bibitem[{\citenamefont{Helstrom}(1976)}]{helstrom1976}
\bibinfo{author}{\bibfnamefont{C.~W.} \bibnamefont{Helstrom}},
  \emph{\bibinfo{title}{Quantum Detection and Estimation}}
  (\bibinfo{publisher}{Academic Press}, \bibinfo{year}{1976}).

\bibitem[{\citenamefont{Matsumoto}(2002)}]{matsumoto2002}
\bibinfo{author}{\bibfnamefont{K.}~\bibnamefont{Matsumoto}},
  \bibinfo{journal}{J. Phys. A} \textbf{\bibinfo{volume}{35}},
  \bibinfo{pages}{3111} (\bibinfo{year}{2002}).

\bibitem[{\citenamefont{Majorana}(1932)}]{Majorana1932}
\bibinfo{author}{\bibfnamefont{E.}~\bibnamefont{Majorana}},
  \bibinfo{journal}{Nuovo Cimento} \textbf{\bibinfo{volume}{9}},
  \bibinfo{pages}{43} (\bibinfo{year}{1932}).

\bibitem[{\citenamefont{Penrose and Rindler}(1984)}]{Prenrose1984}
\bibinfo{author}{\bibfnamefont{G.}~\bibnamefont{Penrose}} \bibnamefont{and}
  \bibinfo{author}{\bibfnamefont{W.}~\bibnamefont{Rindler}},
  \emph{\bibinfo{title}{Spinors and Space-Time}} (\bibinfo{publisher}{Cambridge
  Univeristy Press}, \bibinfo{year}{1984}).

\bibitem[{\citenamefont{Zimba}(2006)}]{Zimba2006}
\bibinfo{author}{\bibfnamefont{J.}~\bibnamefont{Zimba}},
  \bibinfo{journal}{Electr. J. Theor. Phys} \textbf{\bibinfo{volume}{3}},
  \bibinfo{pages}{143} (\bibinfo{year}{2006}).

\bibitem[{\citenamefont{Ballester}(2004)}]{ballester2004}
\bibinfo{author}{\bibfnamefont{M.~A.} \bibnamefont{Ballester}},
  \bibinfo{journal}{Phys. Rev. A} \textbf{\bibinfo{volume}{69}},
  \bibinfo{pages}{022303} (\bibinfo{year}{2004}).

\bibitem[{\citenamefont{Ballester}(2005)}]{ballester2005}
\bibinfo{author}{\bibfnamefont{M.~A.} \bibnamefont{Ballester}},
  \bibinfo{journal}{arxiv:quant-ph/0507073}  (\bibinfo{year}{2005}).

\bibitem[{\citenamefont{Berry and Wiseman}(2000)}]{berry2000}
\bibinfo{author}{\bibfnamefont{D.~W.} \bibnamefont{Berry}} \bibnamefont{and}
  \bibinfo{author}{\bibfnamefont{H.~M.} \bibnamefont{Wiseman}},
  \bibinfo{journal}{Phys. Rev. Lett.} \textbf{\bibinfo{volume}{85}},
  \bibinfo{pages}{5098} (\bibinfo{year}{2000}).

\bibitem[{\citenamefont{Bagan et~al.}(2001)\citenamefont{Bagan, Baig, Brey,
  Munoz-Tapia, and Tarrach}}]{bagan2001}
\bibinfo{author}{\bibfnamefont{E.}~\bibnamefont{Bagan}},
  \bibinfo{author}{\bibfnamefont{M.}~\bibnamefont{Baig}},
  \bibinfo{author}{\bibfnamefont{A.}~\bibnamefont{Brey}},
  \bibinfo{author}{\bibfnamefont{R.}~\bibnamefont{Munoz-Tapia}},
  \bibnamefont{and} \bibinfo{author}{\bibfnamefont{R.}~\bibnamefont{Tarrach}},
  \bibinfo{journal}{Phys. Rev. A} \textbf{\bibinfo{volume}{63}},
  \bibinfo{pages}{052309} (\bibinfo{year}{2001}).

\bibitem[{\citenamefont{Chiribella et~al.}(2004)\citenamefont{Chiribella,
  DAriano, Perinotti, and Sacchi}}]{Chiribella2004}
\bibinfo{author}{\bibfnamefont{G.}~\bibnamefont{Chiribella}},
  \bibinfo{author}{\bibfnamefont{G.~M.} \bibnamefont{DAriano}},
  \bibinfo{author}{\bibfnamefont{P.}~\bibnamefont{Perinotti}},
  \bibnamefont{and} \bibinfo{author}{\bibfnamefont{M.~F.}
  \bibnamefont{Sacchi}}, \bibinfo{journal}{Phys. Rev. Lett.}
  \textbf{\bibinfo{volume}{93}}, \bibinfo{pages}{180503}
  (\bibinfo{year}{2004}).

\bibitem[{\citenamefont{Bollinger et~al.}(1996)\citenamefont{Bollinger, Itano,
  Wineland, and Heinzen}}]{bollinger1996}
\bibinfo{author}{\bibfnamefont{J.~J.~.} \bibnamefont{Bollinger}},
  \bibinfo{author}{\bibfnamefont{W.~M.} \bibnamefont{Itano}},
  \bibinfo{author}{\bibfnamefont{D.~J.} \bibnamefont{Wineland}},
  \bibnamefont{and} \bibinfo{author}{\bibfnamefont{D.~J.}
  \bibnamefont{Heinzen}}, \bibinfo{journal}{Phys. Rev. A}
  \textbf{\bibinfo{volume}{54}}, \bibinfo{pages}{R4649} (\bibinfo{year}{1996}).

\bibitem[{\citenamefont{Holevo}(1982)}]{holevo1982}
\bibinfo{author}{\bibfnamefont{A.~S.} \bibnamefont{Holevo}},
  \emph{\bibinfo{title}{Probabilistic and Stataistical Aspects of Quantum
  Theory}} (\bibinfo{publisher}{North Holland, Amsterdam},
  \bibinfo{year}{1982}).

\end{thebibliography}

\end{document}